\documentclass[prb,superscriptaddress,twocolumn,showpacs,preprintnumbers,amsmath,floatfix]{revtex4}
\usepackage{amssymb}
\usepackage{graphicx}

\begin{document}

\title{A General Mechanism for Orbital Selective Phase Transitions}

\author{Yu-Zhong Zhang}
\email[Corresponding author. ]{Email: yzzhang@tongji.edu.cn}
\affiliation{Shanghai Key Laboratory of Special Artificial Microstructure Materials and Technology, \\
$\&$ Physics Department, Tongji University, Shanghai 200092, P.R. China}
\author{Hunpyo Lee}
\affiliation{Institut f\"ur Theoretische Physik, Goethe-Universit\"at Frankfurt, Max-von-Laue-Strasse 1, 60438 Frankfurt/Main, Germany}
\author{Hai-Qing Lin}
\affiliation{Beijing Computational Science Research Center, Beijing 100084, China}
\author{Chang-Qin Wu}
\affiliation{Department of Physics and State Key Laboratory of Surface Physics, Fudan University, Shanghai 200433, China}
\author{Harald O. Jeschke}
\affiliation{Institut f\"ur Theoretische Physik, Goethe-Universit\"at Frankfurt, Max-von-Laue-Strasse 1, 60438 Frankfurt/Main, Germany}
\author{Roser Valent{\'\i}}
\affiliation{Institut f\"ur Theoretische Physik, Goethe-Universit\"at Frankfurt, Max-von-Laue-Strasse 1, 60438 Frankfurt/Main, Germany}

\date{\today}

\begin{abstract}

Based on the analysis of a two-orbital Hubbard model within a
mean-field approach, we propose a mechanism for an orbital selective
phase transition (OSPT) where coexistence of localized and itinerant
electrons can be realized. We show that this OSPT exists both at and
near half filling even in the absence of crystal field splittings or
when bandwidths, orbital degeneracies and magnetic states are equal
for both orbitals provided the orbitals have different band
dispersions. Such conditions should be generally satisfied in many
materials. We find that this OSPT is not sensitive to the strength of
Hund's rule coupling and that heavy doping favors the collinear
antiferromagnetic state over the OSPT. We discuss our results in
relation to the iron pnictides.

\end{abstract}

\pacs{71.10.Fd,71.30.+h,71.10.Hf,75.10.-b}

\maketitle

\section{Introduction}
\label{sec:one}
Orbital selective phase transitions (OSPTs) leading to phases
where localized and itinerant electrons coexist have attracted
extensive interest from both
experimentalists~\cite{Wang2004,Balicas2005,Lee2006,Kim2006,Shimoyamada2009,Neupane2009}
and
theoreticians~\cite{Anisimov2002,Koga2004,Knecht05,Koga05,Arita05,Song05,Ferrero05,Medici05,Biermann2005,Liebsch2005,Inaba2006,Costi2007,Bouadim09,Jakobi09,Song09,Lee2010PRL,Medici2011,Kita2011}
since the observation in the metallic phase of Ca$_{2-x}$Sr$_x$RuO$_4$
($0.2 \leq x \leq 0.5$) of an anomalous behavior with a
Curie-Weiss-like local spin~\cite{Nakatsuji2000}.  In spite of the
controversies regarding the applicability of such a proposal to real
compounds~\cite{Fang2002,Ishida2007,Gorelov2010}, various mechanisms
for OSPTs have been investigated, such as two orbitals with different
bandwidth at
half-filling~\cite{Anisimov2002,Koga2004,Knecht05,Koga05,Arita05,Song05,Ferrero05,Medici05,Biermann2005,Liebsch2005,Inaba2006,Costi2007,Bouadim09,Jakobi09,Song09,Lee2010PRL},
away from half filling with crystal field splitting~\cite{Werner2007},
coexistence of different orbital degeneracies with crystal field
splitting at any filling~\cite{Medici2009}, or different magnetic
states in different orbitals at half-filling~\cite{Lee2011}.

Recently, various models based on an assumption of the
coexistence of localized and itinerant electrons have been proposed
in order to describe the magnetism in the new iron-based
superconductors~\cite{Yin2011,You2011,Hackl2009,Yang2010,You2011PRB,Lv2010}. Less work
has been done on understanding the origin of such orbital selective
phases (OSPs).  As one possible mechanism for the OSPTs in the pnictides,
bands with similar bandwidths having different intra-band Coulomb
repulsion were suggested in analogy to the mechanism of a difference
in bandwidth~\cite{Wu2008}.  However, since most of the previous
studies about the origin of OSPs are focused on the paramagnetic
(PM) state, a correct description of the magnetism observed in the
parent compounds of most iron pnictides calls for a re-investigation
of the mechanism responsible for OSPs with magnetic order.

In this paper, a possible OSPT mechanism is proposed based on a
simple two-dimensional (2D) two-orbital Hubbard model with both
orbitals having different band dispersions. We solve the model in the
context of mean-field theory (Hartree-Fock approximation (HFA)).  A
comparison of our results to those obtained from the dynamical
mean-field approximation (DMFA) at and near half filling shows that OSPT can be qualitatively
captured already at the mean-field level without taking dynamical
fluctuations into account. The advantage of working with the
mean-field approach is that we will be able to investigate a large
variety of possible cases not easily accessible within the DMFA.  We
will show that even in the absence of crystal field splittings or when
bandwidths, orbital degeneracies, magnetic states and intra-band
Coulomb repulsion are equal for both orbitals, OSPTs can still occur
at different band fillings.  We will show that it is the distinct band
dispersion in both orbitals that can be identified as the crucial
ingredient for the presence of OSPTs with magnetic order. The
mechanism we consider is in fact very general since usually the
strength of hybridizations between neighboring sites in different
directions is strongly orbital-dependent in real materials, leading to
distinct band dispersions in different orbitals.

\section{Model and Method}
\label{sec:two}
The 2D two-orbital Hubbard model is defined as
\begin{eqnarray}
&H&=-\sum_{\langle ij\rangle, \langle\langle ij\rangle\rangle, \gamma \sigma} t_{ij,\gamma} c^{\dagger}_{i\gamma\sigma}c_{j\gamma\sigma}
+U\sum_{i\gamma}n_{i\gamma\uparrow}n_{i\gamma\downarrow} \nonumber \\
&+&\Big(U'-\frac J 2\Big)\sum_{i\gamma>\gamma'}n_{i\gamma} n_{i\gamma'}-2J\sum_{i\gamma>\gamma'}S_{i\gamma}\cdot S_{i\gamma'},\label{eq:hamiltonian}
\end{eqnarray}
where $t_{ij,\gamma}=t_{\gamma}$ ($t_{\gamma}^{\prime }$) is the
intra-orbital hopping integral between NN (NNN) sites denoted by
$\langle ij\rangle$ ($\langle\langle ij\rangle\rangle$) with band
indices $\gamma=\alpha,\beta$ in units of $t$. $U$, $U^{\prime }$ and
$J$ are the intra-band, inter-band Coulomb interaction and Hund's
coupling, respectively, which fulfill the rotational invariance
condition $U=U^{\prime }+2J$. The pair-hopping term is ignored as it
does not affect our mean-field
results~\cite{Lorenzana2008,Daghofer2008,Kaneshita2009,Bascones2010}. $c^{\dagger}_{i\gamma\sigma}$
($c_{i\gamma\sigma}$) creates (annihilates) an electron in orbital
$\gamma$ of site $i$ with spin $\sigma$. $n_{i\gamma\sigma}$ is the
occupation operator, while
$n_{i\gamma}=n_{i\gamma\uparrow}+n_{i\gamma\downarrow}$, and
$S_{i\gamma}$ the spin operator.

\begin{figure}[tb]
\includegraphics[width=0.48\textwidth]{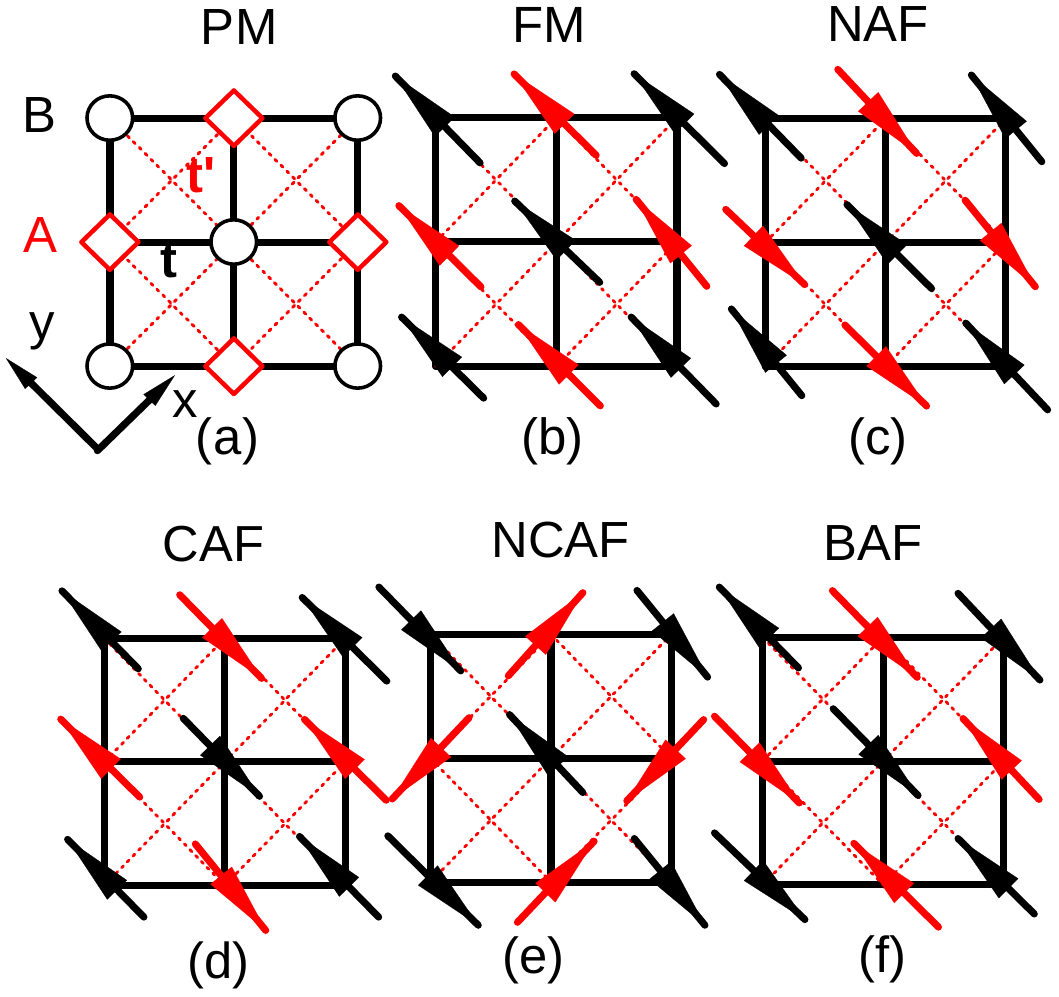}
\caption{(Color online) Cartoons for the different magnetically ordered
  states we use in our calculations. (a) Paramagnetic state. The
  choices of sublattice and coordinate system are shown. (b)
  Ferromagnetic, (c) Neel, (d) collinear, (e)
  noncollinear, (f) bi-collinear antiferromagnetic states.} \label{fig:pattern}
\end{figure}

In order to access the true ground state in a 2D system with hoppings
up to NNN sites, the original lattice is divided into two sublattices
A and B (see Fig.~\ref{fig:pattern} (a)), allowing us to consider
various magnetic states in uniform formulation, such as the PM state,
ferromagnetic (FM) state with momentum $Q_{A/B,\gamma}=(0,0)$ and
magnetization $m^{x/y}_{A,\gamma}=m^{x/y}_{B,\gamma}$, NAF state with
$Q_{A/B,\gamma}=(0,0)$ and $m^{x/y}_{A,\gamma}=-m^{x/y}_{B,\gamma}$,
collinear AF (CAF) state with $Q_{A/B,\gamma}=(\pi,\pi)$ and
$m^{x/y}_{A,\gamma}=m^{x/y}_{B,\gamma}$, bi-collinear AF (BAF) state
with $Q_{A/B,\gamma}=(0,\pi)$ and
$m^{x/y}_{A,\gamma}=m^{x/y}_{B,\gamma}$, and non-collinear AF (NCAF)
state with $Q_{A/B,\gamma}=(\pi,\pi)$ and
$m^{x/y}_{A,\gamma}=m^{y/x}_{B,\gamma}$~\cite{Lorenzana2008}, where
$|m_{i,\gamma}|=|m^{x}_{i,\gamma}+im^{y}_{i,\gamma}|=|\frac 1 N
\sum_{k}\langle
c^{\dagger}_{ki\gamma\uparrow}c_{k+Qi\gamma\downarrow}\rangle|$ with
i=A or B. The corresponding cartoons for different magnetic patterns
are shown in Fig.~\ref{fig:pattern}~(a)-(f).

\section{Comparisons between the results from HFA and DMFA}
\label{sec:three}

\begin{figure}[tb]
\includegraphics[width=0.48\textwidth]{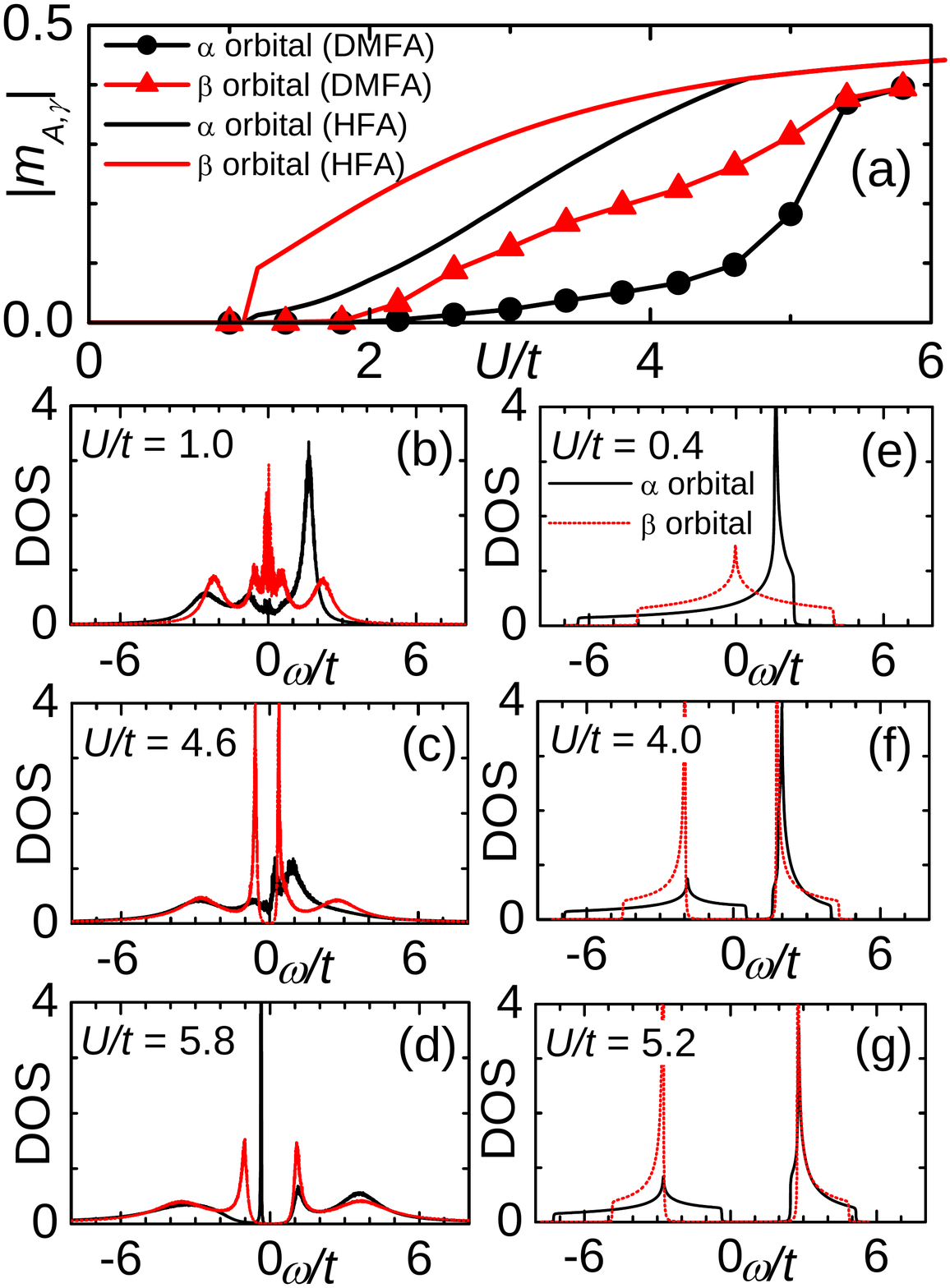}
\caption{(Color online) Comparison of the results from Hartree-Fock
  approximation (HFA) and dynamical mean-field approximation (DMFA) at
  $t_\alpha=1$, $t_\alpha'=0.6$, $t_\beta=1$, $t_\beta'=0$, and
  $J/U=0.25$. a) Magnetization as a function of $U/t$. (b)-(d)
  ((e)-(g)) show the density of states in different phases from DMFA
  (HFA). } \label{fig:comparison}
\end{figure}

In order to check the validity of our mean-field calculations, we
first compare our results with those obtained using the
DMFA~\cite{Georges1996,Kotliar2006,Gull2011}. For this comparison, the chemical potential
rather than the filling is fixed as is usually done in DMFA studies,
and only the NAF state is allowed as required by a two-sublattice
calculation within the DMFA~\cite{Lee2010PRB}. Fig.~\ref{fig:comparison}~(a) shows the
sublattice magnetization as a function of interaction $U/t$ for the
case $t_\alpha=1$, $t_\alpha'=0.6$ and $t_\beta=1$, $t_\beta'=0$. We
find that while the magnetic phase transition obtained from the HFA
happens earlier than that from the DMFA and higher magnetization is
detected in the HFA -indicating that dynamical fluctuations ignored in
the HFA strongly suppress the magnetically ordered states- the
variation of the magnetization with $U/t$ obtained from the DMFA can
be qualitatively reproduced by the results from the HFA. Furthermore,
all the phases given from the DMFA can be qualitatively captured by
the HFA as shown in Figs.~\ref{fig:comparison} (b)-(d) and (e)-(g)
which depict the density of states (DOS) in the different phases
obtained from DMFA and HFA, respectively. The OSPT, where one orbital
becomes localized while the other remains metallic, is clearly
detected by both DMFA (Fig.~\ref{fig:comparison}~(c)) and HFA
(Fig.~\ref{fig:comparison}~(f)). The resulting OSP is sandwiched
between the PM metallic state and the NAF insulating state as seen in
Fig.~\ref{fig:comparison}~(b), (d) for DMFA and
Fig.~\ref{fig:comparison}~(e), (g) for HFA. The qualitative
consistency between the results from HFA and DMFA imply the validity
of our following discussion on the OSP as well as on other phases in
our model at the mean-field level.  In fact, it is already known from
the DMFA -where spatial fluctuations are absent- that the PM
metal-insulator transition which is inaccessible to the HFA is
precluded by a magnetic phase transition in the half-filled case at
zero temperature~\cite{Peters2009} which may be qualitatively
described by the HFA. Comparing DMFA and HFA at other hopping
parameters with different $t_\alpha'$ (not shown here), we concluded
that dynamical fluctuations play a minor role in the OSPT.

\section{The general mechanism for OSPT}
\label{sec:four}

\begin{figure*}[tb]
\includegraphics[width=0.92\textwidth]{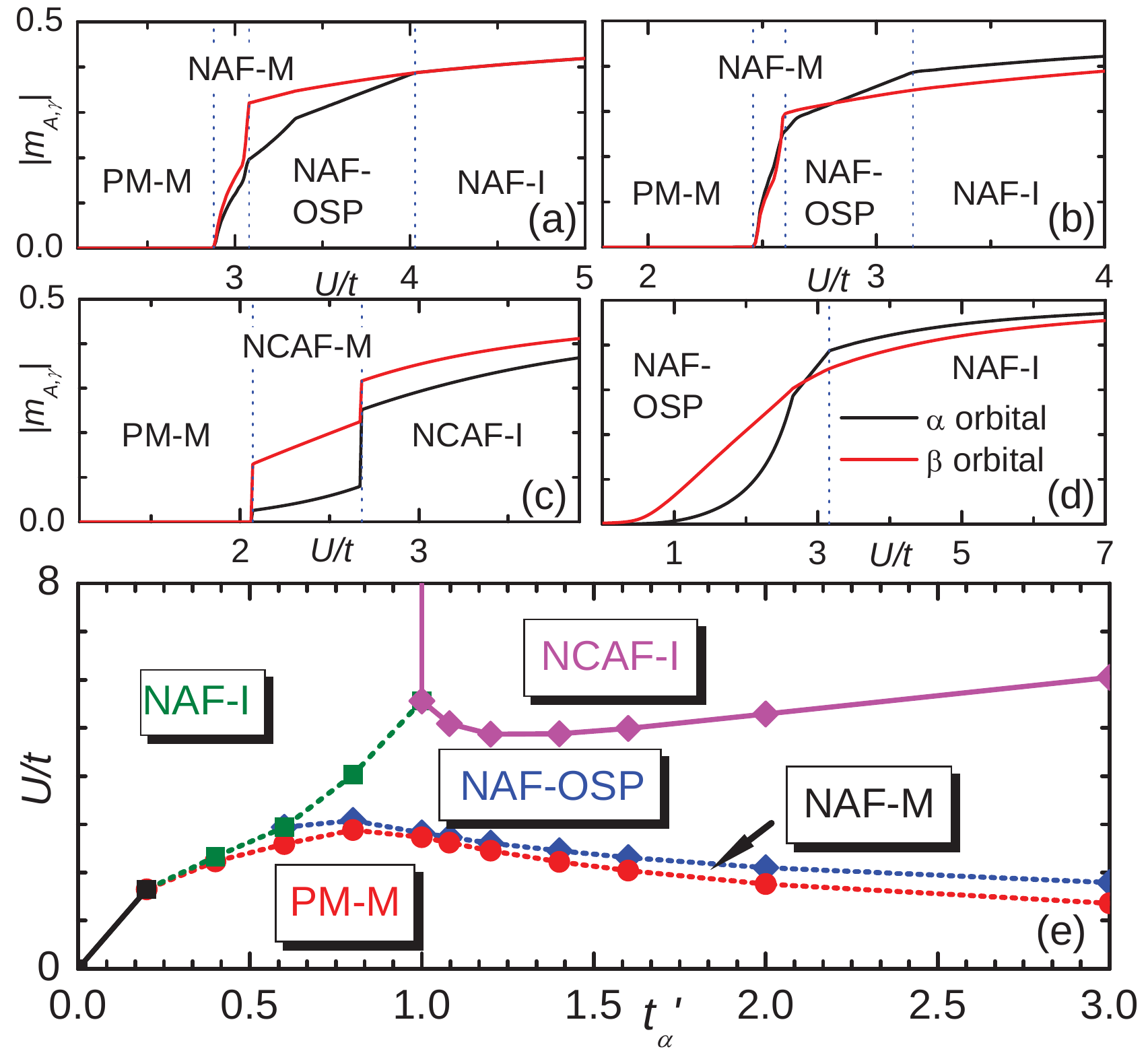}
  \begin{center}
\caption{(Color online) Variation of the magnetization as a function
  of $U/t$ (a) at $t_\alpha=1$, $t_\alpha'=0.8$ and $t_\beta=1$,
  $t_\beta'=0$, (b) at $t_\alpha=0.769$, $t_\alpha'=0.615$ and
  $t_\beta=1$, $t_\beta'=0$, (c) at $t_\alpha=1$, $t_\alpha'=0.8$ and
  $t_\beta=0.769$, $t_\beta'=0.615$, (d) after eliminating the orbital
  order by adding an effective crystal field splitting of
  $\Delta=0.798$ at $t_\alpha=0.769$, $t_\alpha'=0.615$ and
  $t_\beta=1$, $t_\beta'=0$, where the term for crystal field
  splitting is written as $\sum_{i} \Delta
  (n_{i\beta}-n_{i\alpha})$. (e) Phase diagram in $U/t$-$t_\alpha'$ at
  $t_\alpha=1$, $t_\beta=1$, $t_\beta'=0$. Here $J/U=0.25$ and filling
  is $1/2$. Regions of different phases are indicated by the
  abbreviations defined in the text. M (I) denotes metal
  (insulator). Solid and dotted lines represent first and second order phase transitions, respectively.} \label{fig:PD}
  \end{center}
\end{figure*}

In the following we shall investigate the case of fixed filling at
$1/2$ in contrast to the case of fixed chemical potential where the
filling is changed as a function of interaction $U/t$. All the
magnetically ordered states shown in Fig.~\ref{fig:pattern} are taken
into account and the ground state is the one with lowest total
energy. Fig.~\ref{fig:PD}~(a) shows the phase transitions happening at
$t_\alpha=1$, $t_\alpha'=0.8$ and $t_\beta=1$, $t_\beta'=0$ as a
function of $U/t$. As long as $U/t<2.88$, the ground state is a PM
metal with orbital order. In a small interaction region of
$2.88<U/t<3.08$, an NAF metal with orbital order appears. Further
increasing $U/t$ from $3.08$ up to $4.02$, the $\alpha$ orbital
exhibits NAF insulating behavior while the $\beta$ orbital keeps the
NAF metallic state, indicating an OSP. Orbital order disappears in
this interaction region. At $U/t>4.02$, both orbitals display NAF
insulating behavior.

Though we have demonstrated that the OSPT is still present at fixed
filling in finite dimension, the mechanism for it has not yet been
identified. After analyzing the noninteracting DOS, we find that
several possible mechanisms coexist, such as (i) two orbitals having
different bandwidth with the ratio of $W_\alpha/W_\beta=1.3$, (ii) the
existence of orbital order due to the different band dispersions of
the two orbitals ($t_\alpha'/t_\alpha \neq t_\beta'/t_\beta$) which
can be viewed as the existence of an effective crystal field
splitting, and (iii) two orbitals having distinct band dispersions
which leads to different shapes of the noninteracting partial DOS. The
last effect was not considered in previous DMFA studies where a
semicircular DOS with particle-hole symmetry is usually employed for
all the orbitals. In the following, we will reveal that orbitals
having distinct band dispersions play a crucial role in the OSPT with
magnetic order.

In order to figure out the essential mechanism responsible for the
OSPT observed above, we will study separately three cases.  (i) We
first eliminate the effect of different bandwidths by rescaling the
hopping parameters of the $\alpha$ orbital from $t_\alpha=1$,
$t_\alpha'=0.8$ to $t_\alpha=0.769$, $t_\alpha'=0.615$ so that the
ratio of $t_\alpha'/t_\alpha=0.8$ is retained while the ratio of
bandwidths becomes $W_\alpha/W_\beta=1$. Fig.~\ref{fig:PD}~(b)
presents the various phases as a function of $U/t$
after rescaling. Though the critical points are changed due to the
change of the total bandwidths, all the phases involving OSP are
preserved, indicating that such an OSPT exists in the absence of bandwidth
differences between orbitals.

As a second case (ii), we remove the orbital order by adding an
effective crystal field splitting, by which the half-filling condition
is simultaneously satisfied at $U/t=0$ in both orbitals.
Fig.~\ref{fig:PD}~(d) shows that the OSPT is still present in the absence
of orbital order. However, the states with metallic behavior in both
orbitals vanish since the Fermi level is located right
at the van Hove singularity in the $\beta$ orbital at $U/t=0$. We have
checked that a small $t_\beta'$ which shifts the van Hove singularity
away from the Fermi level leads to the appearance of metallic phases
in both orbitals at finite $U/t$.

As a third option (iii), we eliminate the effect of orbitals having
distinct band dispersions but retain the difference in bandwidth by
choosing $t_\alpha=1$, $t_\alpha'=0.8$ and $t_\beta=0.769$,
$t_\beta'=0.615$ which leads to
$t_\alpha'/t_\alpha=t_\beta'/t_\beta=0.8$ and
$W_\alpha/W_\beta=1.3$. As shown in Fig.~\ref{fig:PD}~(c), an OSP is
precluded by NCAF states, resulting in only two successive phase
transitions from PM metals to NCAF insulators through NCAF metals in
both orbitals. Clearly, the OSP will be replaced by NAF insulating
states in both orbitals at any finite $U/t$ if we take $t_\alpha=1$,
$t_\alpha'=0$ and $t_\beta=1.3$, $t_\beta'=0$, which means a similar
dispersion relation $t_\alpha'/t_\alpha=t_\beta'/t_\beta=0$ but
different bandwidth $W_\alpha/W_\beta=1.3$, since the Fermi level
crosses the van Hove singularities in both orbitals.

Our results so far strongly point to the fact that orbitals with
distinct band dispersions are crucial for the OSPT since it is present
even though all the other mechanisms mentioned above are
absent while different bandwidth alone will not support the existence
of OSPT when magnetic order is considered. Fig.~\ref{fig:PD}~(e)
presents a phase diagram in the $U/t$-$t_\alpha'$ plane at $t_\alpha=1$,
$t_\beta=1$, and $t_\beta'=0$. An OSP exists in a
wide region of the phase diagram. The phase transitions from both NAF
states to the NCAF state and from PM metal to NAF insulator are of
first order (solid line), otherwise second order (dotted line). The
NAF metallic state has also been detected in the one-band Hubbard
model with NN and NNN hoppings~\cite{Yu2010}.

\section{Various effects on the observed OSPT}
\label{sec:five}

\begin{figure*}[tb]
  \begin{center}
\includegraphics[width=0.92\textwidth]{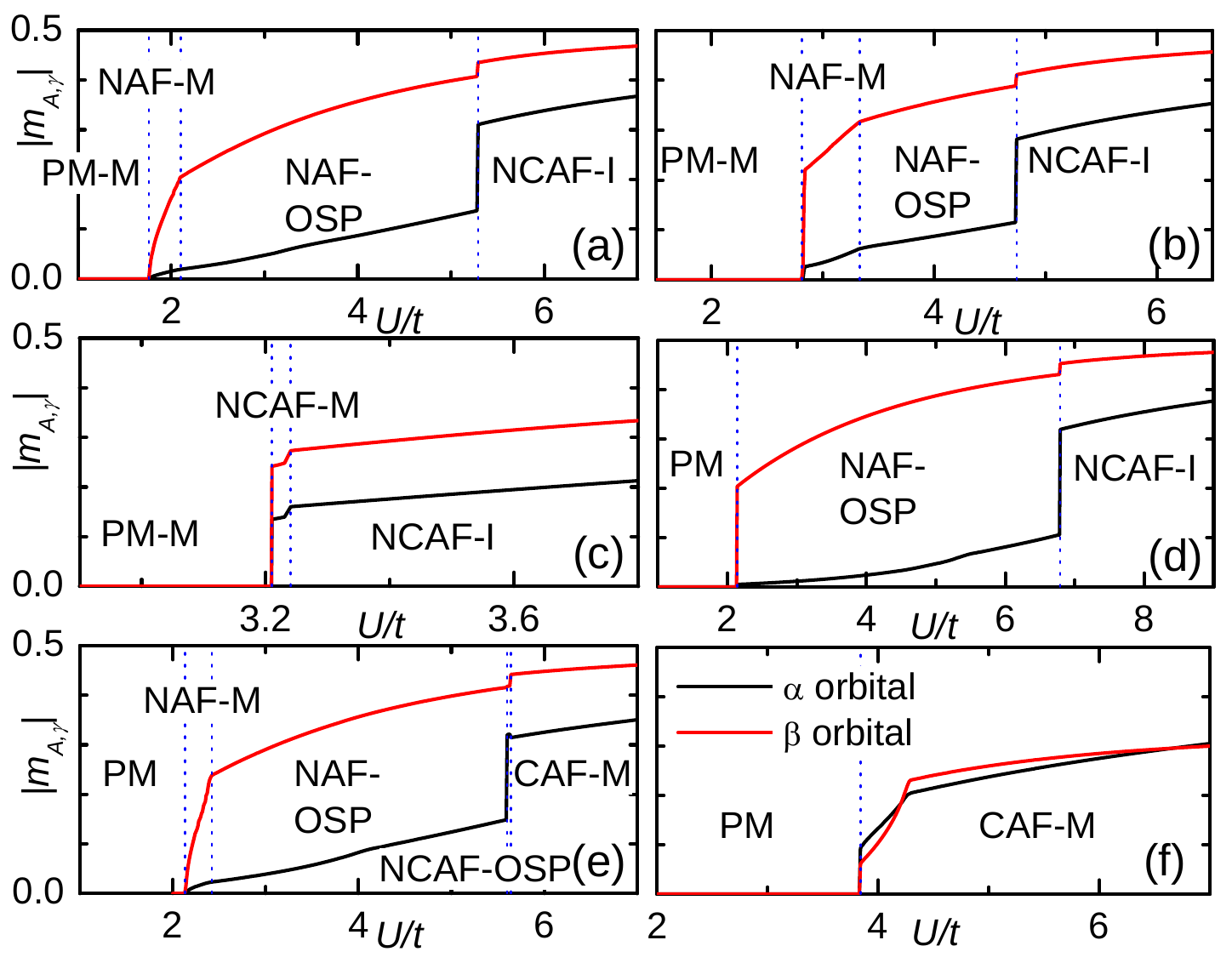}
\caption{(Color online) Variation of the magnetization as a function of
  $U/t$ at $t_\alpha=1$, $t_\alpha'=2$, $t_\beta=1$. (a) $t_\beta'=0$,
  $J/U=0.25$, (b) $t_\beta'=0.4$, $J/U=0.25$, (c) $t_\beta'=0.8$,
  $J/U=0.25$, and (d) $t_\beta'=0$, $J/U=0.0625$ at half-filling. (e)
  $2.5\%$ and (f) $20\%$ electronic doping at $t_\beta'=0$,
  $J/U=0.25$. Regions of different phases are indicated by the
  abbreviations defined in the text. M (I) denotes metal
  (insulator).} \label{fig:othereffects}
  \end{center}
\end{figure*}

Finally, we investigate various effects on the observed OSPT. From the
phase diagram, it is obvious that we should discuss two cases
separately: 1) $t_\alpha'/t_\alpha>1$ where different magnetic orders
like NAF and NCAF orders compete with each other; 2)
$t_\alpha'/t_\alpha<1$ where only NAF order occurs. In
Fig.~\ref{fig:othereffects}, we show the results at $t_\alpha=1$,
$t_\beta=1$ and $t_\alpha'=2$. We first present the effect of adding
NNN hopping $t_\beta'$. It is found that increasing $t_\beta'$ favors
the NCAF state, which squeezes the region of the NAF OSP. As seen in
Fig.~\ref{fig:othereffects}~(b), at $t_\beta'=0.4$, the region of OSP
is smaller than that at $t_\beta'=0$ (Fig.~\ref{fig:othereffects}~(a))
and at $t_\beta'=0.8$ the OSP completely vanishes (see
Fig.~\ref{fig:othereffects}~(c)). However, for the case of
$t_\alpha'=0.8$ (not shown), the region of the OSP remains unchanged
at $t_\beta'=0.4$, while it is reasonably replaced by NCAF state at
$t_\beta'=0.8$. The effect of Hund's rule coupling is presented in
Fig.~\ref{fig:othereffects}~(d). Compared to
Fig.~\ref{fig:othereffects}~(a) where $J/U=0.25$, the region of OSP is
enlarged at $J/U=0.0625$ and a direct first-order phase transition
from PM metals in both orbitals to the NAF OSP is observed instead of
two successive second-order phase transitions through an intermediate
NAF metallic state at $J/U=0.25$. For the case of $t_\alpha'=0.8$,
$t_\beta'=0$ (not shown), a similar effect of the Hund's rule coupling
is found.

Fig.~\ref{fig:othereffects}~(e) shows that at a small concentration of
electronic doping of $2.5\%$, the OSP with NAF order is slightly moved
to higher values of $U/t$ and the NCAF insulating states existing in
the undoped case is replaced by a small region of OSP with NCAF order
which eventually become CAF metallic states at larger $U/t$. At large
doping of $20\%$, only two phases with PM and CAF metallic states
remain and the OSP vanishes as seen in
Fig.~\ref{fig:othereffects}~(f). The critical value of doping
concentration where the OSP disappears is around $13.6\%$. For the
case of $t_\alpha'=0.8$, $t_\beta'=0$ (not shown), the OSPT also
exists at $2.5\%$ doping but is excluded by PM and CAF metallic
states at $20\%$ doping. It is interesting to notice that CAF metallic
states only appear when the system is doped. $20\%$ electronic doping
is related to the filling factor in the pnictides where $6$ 3d
electrons occupy $5$ 3d orbitals. However, after examining various
sets of model parameters, including those for the pnictides~\cite{Daghofer2008,Raghu2008}, we should emphasize that the OSPT disappears
whenever CAF order occurs.

\section{Discussions and conclusions}
\label{sec:six}
Recently, various efforts have been made in order to reconcile the
controversies about the origin of the CAF phases observed in the iron
pnictides. Models containing coupled local spins and itinerant
electrons have been
proposed~\cite{Yin2011,Hackl2009,Yang2010,You2011PRB,Lv2010,Akbari2011}. Experimental data has been
also interpreted in terms of a coexistence of local and itinerant
electrons~\cite{Zhang2010,Yuan2011}. However, such a compromise
doesn't seem to be supported by the present study of OSPT with
magnetic order. We find that OSPT and CAF order tend to avoid each
other. Also, involving the inter-orbital hoppings do not favor the OSP with CAF order. On the other hand, it is not to be expected that increasing the orbital
degrees of freedom will dramatically change the situation. Furthermore, existing mechanisms
proposed within the PM state are in conflict with the fact that the
low temperature phases of most pnictides are magnetically ordered,
and the bandwidth of different orbitals are almost the same. However,
quantum fluctuations, especially spatial rather than dynamical
fluctuations, which favors paramagnetic states, may be
responsible for possible OSPT in the pnictides.

In summary, we propose a general mechanism for an OSPT in magnetically
ordered states. Different orbitals with different band dispersions
should be quite widespread in real materials.
Importantly, the OSPT according to the presented mechanism occurs in a
wide range of model parameters, suggesting that this mechanism could
be realized in nature.

{\it Acknowledgments.-} YZ is supported by National Natural Science
Foundation of China (No. 11174219), Shanghai Pujiang Program
(No. 11PJ1409900) and Research Fund for the Doctoral Program of Higher Education of China (No. 20110072110044). HL is supported by the DFG through FOR 1346, and
HOJ by the Helmholtz Association through grant HA216/EMMI. YZ is
indebted to CSRC for the hospitality and partial financial support
from CAEP.


\begin{thebibliography}{99}

\bibitem{Wang2004} S.-C. Wang, H.-B. Yang, A. K. P. Sekharan, S. Souma, H. Matsui, T. Sato, T. Takahashi, Chenxi Lu, Jiandi Zhang, R. Jin, D. Mandrus, E. W. Plummer, Z. Wang, and H. Ding, Phys. Rev. Lett. \textbf{93}, 177007 (2004).

\bibitem{Balicas2005} L. Balicas, S. Nakatsuji, D. Hall, T. Ohnishi, Z. Fisk, Y. Maeno, and D. J. Singh, Phys. Rev. Lett. \textbf{95}, 196407 (2005).

\bibitem{Lee2006} J. S. Lee, S. J. Moon, T. W. Noh, S. Nakatsuji, and Y. Maeno, Phys. Rev. Lett. \textbf{96}, 057401 (2006).

\bibitem{Kim2006} B. J. Kim, J. Yu, H. Koh, I. Nagai, S. I. Ikeda, S.-J. Oh, and C. Kim, Phys. Rev. Lett. \textbf{97}, 106401 (2006).

\bibitem{Shimoyamada2009} A. Shimoyamada, K. Ishizaka, S. Tsuda, S. Nakatsuji, Y. Maeno, and S. Shin, Phys. Rev. Lett. \textbf{102}, 086401 (2009).

\bibitem{Neupane2009} M. Neupane, P. Richard, Z.-H. Pan, Y.-M. Xu, R. Jin, D. Mandrus, X. Dai, Z. Fang, Z. Wang, and H. Ding, Phys. Rev. Lett. \textbf{103}, 097001 (2009).

\bibitem{Anisimov2002} V. I. Anisimov, I. A. Nekrasov, D. E. Kondakov,  T. M. Rice, and M. Sigrist, Eur. Phys. J. B \textbf{25}, 191 (2002).

\bibitem{Koga2004} A. Koga, N. Kawakami, T. M. Rice, and M. Sigrist, Phys. Rev. Lett.  \textbf{92}, 216402 (2004).

\bibitem{Knecht05} C. Knecht, N. Bl{\"u}mer, and P. G. J. van Dongen, Phys. Rev. B \textbf{72}, 081103 (2005).

\bibitem{Koga05} A. Koga, N. Kawakami, T. M. Rice, and M. Sigrist, Phys. Rev. B \textbf{72}, 045128 (2005).

\bibitem{Arita05} R. Arita and K. Held, Phys. Rev. B \textbf{72}, 201102 (2005).

\bibitem{Song05} Y. Song and L.-J. Zou, Phys. Rev. B \textbf{72}, 085114 (2005).

\bibitem{Ferrero05} M. Ferrero, F. Becca, M. Fabrizio, and M. Capone, Phys. Rev. B \textbf{72}, 205126 (2005).

\bibitem{Medici05} L. de{'} Medici, A. Georges, and S. Biermann, Phys. Rev. B \textbf{72}, 205124 (2005).

\bibitem{Biermann2005} S. Biermann, L. de{'} Medici, and A. Georges, Phys. Rev. Lett. \textbf{95}, 206401 (2005).

\bibitem{Liebsch2005} A. Liebsch, Phys. Rev. Lett. \textbf{95}, 116402 (2005).

\bibitem{Inaba2006} K. Inaba and A. Koga, Phys. Rev. B \textbf{73}, 155106 (2006).

\bibitem{Costi2007} T. A. Costi and A. Liebsch, Phys. Rev. Lett. \textbf{99}, 236404 (2007)

\bibitem{Bouadim09} K. Bouadim, G. G. Batrouni, and R. T. Scalettar, Phys. Rev. Lett. \textbf{102}, 226402 (2009).

\bibitem{Jakobi09} E. Jakobi, N. Bl\"umer, and P. van Dongen, Phys. Rev. B \textbf{80}, 115109 (2009).

\bibitem{Song09} Y. Song and L.-J. Zou, Eur. Phys. J. B \textbf{72}, 59 (2009).

\bibitem{Lee2010PRL} H. Lee, Y.-Z. Zhang, H. O. Jeschke, R. Valent{\'\i},  and H. Monien, Phys. Rev. Lett. \textbf{104}, 026402 (2010).

\bibitem{Medici2011} L. de{'} Medici, Phys. Rev. B \textbf{83}, 205112 (2011).

\bibitem{Kita2011} T. Kita, T. Ohashi, and N. Kawakami, Phys. Rev. B \textbf{84}, 195130 (2011).

\bibitem{Nakatsuji2000} S. Nakatsuji, Y. Maeno, Phys. Rev.  Lett. \textbf{84}, 2666 (2000).

\bibitem{Fang2002} Z. Fang, N. Nagaosa, and K. Terakura, Phys. Rev. B 69, 045116 (2004).

\bibitem{Ishida2007} A. Liebsch, H. Ishida, Phys. Rev. Lett. 98, 216403 (2007).

\bibitem{Gorelov2010} E. Gorelov, M. Karolak, T. O. Wehling, F. Lechermann, A. I. Lichtenstein, and E. Pavarini, Phys. Rev. Lett. 104, 226401 (2010).

\bibitem{Werner2007} P. Werner, A. J. Millis, Phys. Rev. Lett. \textbf{99}, 126405 (2007).

\bibitem{Medici2009} L. de{'} Medici, S. R. Hassan, M. Capone, and X. Dai, Phys. Rev. Lett. \textbf{102}, 126401 (2009);L. de{'} Medici, S. R. Hassan and M. Capone, J. Supercond. Nov. Magn. \textbf{22}, 535 (2009).

\bibitem{Lee2011} H. Lee, Y.-Z. Zhang, H. O. Jeschke, and  R. Valent{\'\i}, Phys. Rev. B \textbf{84}, 020401(R). (2011).

\bibitem{Yin2011} W.-G. Yin, C.-C. Lee, and W. Ku, Phys. Rev. Lett. \textbf{105}, 107004 (2010).

\bibitem{You2011} Y.-Z. You, F. Yang, S.-P. Kou, and Z.-Y. Weng, Phys. Rev. Lett. \textbf{107}, 167001 (2011).

\bibitem{Hackl2009} A. Hackl and M. Vojta, New J. Phys. \textbf{11}, 055064 (2009).

\bibitem{Yang2010} F. Yang, S.-P. Kou, and Z.-Y. Weng, Phys. Rev. B \textbf{81}, 245130 (2010).

\bibitem{You2011PRB} Y.-Z. You, F. Yang, S.-P. Kou, Z.-Y. Weng, Phys. Rev. B \textbf{84}, 054527 (2011).

\bibitem{Lv2010} W. Lv, F. Kr{\"u}ger, and P. Phillips, Phys. Rev. B \textbf{82}, 045125 (2010).

\bibitem{Wu2008} J. Wu, P. Phillips, and A. H. Castro Neto, Phys. Rev. Lett. \textbf{101}, 126401 (2008).

\bibitem{Lorenzana2008} J. Lorenzana, G. Seibold, C. Ortix, and M. Grilli, Phys. Rev. Lett. \textbf{101}, 186402 (2008);

\bibitem{Daghofer2008} M. Daghofer, A. Moreo, J. A. Riera, E. Arrigoni, D. J. Scalapino, and E. Dagotto, Phys. Rev. Lett. \textbf{101}, 237004 (2008).

\bibitem{Kaneshita2009} E. Kaneshita, T. Morinari, and T. Tohyama, Phys. Rev. Lett. \textbf{103}, 247202 (2009).

\bibitem{Bascones2010} E. Bascones, M. J. Calder{\'\o}n, and B. Valenzuela, Phys. Rev. Lett. \textbf{104}, 227201 (2010).

\bibitem{Georges1996} A. Georges, G. Kotliar, W. Krauth and M. J. Rozenberg, Rev. Mod. Phys. \textbf{68}, 13 (1996).

\bibitem{Kotliar2006} G. Kotliar, S. Y. Savrasov, K. Haule, V. S. Oudovenko, O. Parcollet, and C. A. Marianetti, Rev. Mod. Phys. \textbf{78}, 865 (2006).

\bibitem{Gull2011} E. Gull, A. J. Millis, A. I. Lichtenstein, A. N. Rubtsov, M. Troyer, and P. Werner, Rev. Mod. Phys. \textbf{83}, 349 (2011).

\bibitem{Lee2010PRB} H. Lee, Y.-Z. Zhang, H. O. Jeschke, and  R. Valent{\'\i}, Phys. Rev. B \textbf{81}, 220506(R) (2010).

\bibitem{Peters2009} R. Peters, T. Pruschke, Phys. Rev. B \textbf{79}, 045108 (2009).

\bibitem{Yu2010} Z.-Q. Yu, L. Yin, Phys. Rev. B \textbf{81}, 195122 (2010).

\bibitem{Raghu2008} S. Raghu, X.-L. Qi, C.-X. Liu, D. J. Scalapino, and S.-C. Zhang, Phys. Rev. B \textbf{77}, 220503 (2008).

\bibitem{Akbari2011} A. Akbari, I. Eremin, and P. Thalmeier, Phys. Rev. B \textbf{84}, 134513 (2011).

\bibitem{Zhang2010} C. Zhang, M. Wang, H. Luo, M. Wang, M. Liu, J. Zhao, D. L. Abernathy, K.Marty, M. D. Lumsden, S. Chi, S. Chang, J. A. Rodriguez-Rivera, J. W. Lynn, T. Xiang, J. Hu, P. Dai, Sci. Rep. \textbf{1}, 115 (2011).

\bibitem{Yuan2011} H. Q. Yuan, L. Jiao, F. F. Balakirev,J. Singleton, C. Setty, J. P. Hu, T. Shang, L. J. Li, G. H. Cao, Z. A. Xu, B. Shen, H. H. Wen, arXiv:1102.5476v1.

\end{thebibliography}
\end{document}